\documentclass{elsart}
\usepackage{graphics}
\usepackage{amssymb}
\usepackage{bm}

\begin{document}
\begin{frontmatter}

\title{Majorization properties of generalized thermal distributions}

\author{N. Canosa, R. Rossignoli, M. Portesi}
\address{Departamento de F\'{\i}sica, Facultad de Ciencias Exactas,
\\ Universidad Nacional de La Plata,
C.C.67, La Plata (1900), Argentina}

\begin{abstract}
We examine the majorization properties of general thermal-like mixed states
depending on a set of parameters. Sufficient conditions which ensure the
increase in mixedness, and hence of {\it any} associated entropic form, when
these parameters are varied, are identified. We then discuss those exhibiting a
power law distribution, showing that they can be characterized by two distinct
mixing parameters, one associated with temperature and the other with the
non-extensivity index $q$. Illustrative numerical results are also provided.
\end{abstract}
\end{frontmatter}

\maketitle

The rigorous concept of disorder derived from the theory of majorization
\cite{HLP.78,MO.79,W.78,B.97} has recently received renewed attention in
theoretical physics, particularly in the field of quantum information
\cite{NC.00,N.99,NK.01,O.04,O.05}. The essential reason is that it is stronger than that
based on standard entropic considerations. The basic idea is that a given
probability distribution or density matrix can be said to be ``more mixed'' or
``disordered'' than another only when it is {\it majorized} by the latter. This
implies a higher entropy of the former, although the converse implication is
not necessarily true. Majorization provides a natural partial ordering on
probability distributions \cite{MO.79} and has consequently found many
applications not only in mathematics, but also in other areas such as economy
and computer science. Moreover, majorization relations are often naturally
satisfied. For instance, in discrete classical systems, the joint distribution
of two random variables is always majorized by the marginal distributions,
while in quantum mechanics, the global density matrix $\rho$ of a {\it
separable} (i.e., non-entangled) mixed state of a composite system is always 
more mixed than  the local reduced densities $\rho_i$ of each subsystem 
\cite{NK.01}, a property which can be violated by entangled states. These 
statements are stronger than the corresponding entropic inequalities [i.e., 
$S(\rho)>S(\rho_i)$].

The aim of this work is to examine along the previous lines the majorization
properties of general thermal-like mixed states depending on a set of
parameters, discussing as application those characterized by a power-law
distribution \cite{TS.88,Cu.91,TM.98}. The latter have in recent years been
analyzed and successfully employed in a wide range of contexts
\cite{T.04,AO.01,TS.04,AR.03,RC.04}, and can be derived within a generalized
nonextensive thermodynamic formalism based on the Tsallis entropy \cite{TS.88}.
One of the basic physical questions we want to answer is if such states {\it do
become more disordered, in the way determined by majorization}, when the
temperature (or some other fundamental parameter characterizing the
distribution) is increased, as occurs with the standard Boltzmann-Gibbs thermal
state. Such property would have far reaching consequences, in particular that
of ensuring a {\it universal} entropy increase, i.e., an increase in {\it any}
consistent disorder measure, and not just in that employed in the construction
of the state. It would also imply the increase of the expectation value of {\it
any} increasing function of energy, and not just of the energy itself. Here we
will prove that such property is indeed valid.

For this purpose, we first identify the sufficient conditions that ensure the
increase in mixedness of a general thermal-like mixed state when the parameters
that characterize it are varied. We then show that states exhibiting a
power-law distribution can be characterized by {\it two} distinct mixing
parameters, one associated with temperature and the other with the
non-additivity index $q$. We also discuss the majorization properties of escort
distributions and the mixing conditions for generalized thermal-like states in 
the presence of constraints on non-commuting observables. Illustrative numerical 
results for a simple model are provided as well. Distributions with correct 
mixing properties can then be employed to investigate the effects of randomness
in physical systems in a more complete way than that provided by the standard
thermal state, allowing to explore different paths from the ground state
to the full random state, all characterized by a proper disorder increase.

 We will consider a quantum system of finite dimension $n$. The
eigenvalues $p_i$ of any density matrix $\rho$ for such system ($p_i\geq 0$,
$\sum_{i=1}^np_i=1$) will be sorted in what follows in {\it decreasing} order
($p_i\geq p_j$ if $i<j$). A density $\rho$ is then said to be {\it more mixed}
than a second density $\rho'$ ($\rho\prec\rho'$) if the eigenvalues of $\rho$
are {\it majorized} by those of $\rho'$ \cite{MO.79,W.78,B.97}:
 \begin{equation}
\rho\prec\rho'\Leftrightarrow s_j\equiv\sum_{i=1}^jp_i\leq
s'_j\equiv\sum_{i=1}^jp'_i\,,\;\;j=1,\ldots,n-1\,,\label{dr}
 \end{equation}
with $s_n=s'_n=1$. In such a case, the probabilities $p_i$ are more ``spread
out'' than the $p'_i$'s, and can be written as a convex combination of
permutations of the latter, i.e., $\mathbf{p}=\sum_\alpha q_\alpha
P_\alpha(\mathbf{p'})$, where $P_\alpha$ are permutations and $q_\alpha\geq 0$,
$\sum_\alpha q_\alpha=1$ \cite{MO.79,W.78,B.97}. The state described by $\rho$
is then more mixed or ``random'' than that described by $\rho'$. Accordingly,
the completely random state $\rho=I/n$ (with $I$ the identity) is more mixed
than any density, while any density is more mixed than a pure state, i.e.,
$I/n\prec\rho\prec |\Phi\rangle\langle\Phi|$ $\forall$ normalized density
$\rho$ and pure state $|\Phi\rangle$.  It can be also shown that if
$\rho\prec\rho'$, $\rho$ can be written as a convex combination of unitary
transformations of $\rho'$, i.e., $\rho=\sum_\alpha q_\alpha
U^\dagger_\alpha\rho' U_\alpha$, with $q_\alpha>0$ and $U_\alpha^\dagger
U_\alpha=I$, and viceversa (Uhlmann's theorem \cite{W.78}). If the dimensions
of $\rho$ and $\rho'$ differ, the same definition (\ref{dr}) can be applied
after completing with zeros the set of eigenvalues of the density of lowest
dimension.

Let us briefly discuss now the relation with entropy. Consider for instance the
general entropic forms \cite{M.97,Pp.97,CR.02,Cu.99}
\begin{equation}
S_f(\rho)={\rm Tr}f(\rho)=\sum_{i=1}^n\,f(p_i)\,,
\end{equation}
where $f$ is a smooth strictly concave function ($f'(p_i)<f'(p_j)$ if
$p_i>p_j$) defined in the interval $[0,1]$, satisfying $f(0)=f(1)=0$. The von
Neumann entropy $S(\rho)=-{\rm Tr}\rho\ln\rho$ and the Tsallis generalization
\cite{TS.88},
\begin{equation}
S_q(\rho)={\rm Tr}(\rho-\rho^q)/(q-1)\,,\;\;q>0\,,
\end{equation}
which approaches the von Neumann entropy for $q\rightarrow 1$, are the most
important examples. It can be shown that if $\rho\prec\rho'\Rightarrow
S_f(\rho)\geq S_f(\rho')$ for {\it any} $f$ of the previous form
\cite{W.78,RC.03} (the same holds for the Renyi entropy
$S_q^R=\ln[1+(1-q)S_q(\rho)]/(1-q)$ \cite{BS.93}, since it is an {\it
increasing} function of $S_q(\rho)$, as well as for any Schur concave function
of $\rho$ \cite{B.97}). However, for a given $f$, the converse is not
necessarily true, so that the concept of disorder implied by Eqs.\ (\ref{dr})
is stronger than that based on a particular choice of $f$. Nonetheless, the
converse holds as follows: if $S_f(\rho)\geq S_f(\rho')$ for {\it any} $f$ of
the previous form $\Rightarrow$ $\rho\prec\rho'$ \cite{W.78,RC.03}. In other
words, the hallmark of increasing mixedness is a {\it universal} entropy
increase. Note, however, that Eqs.\ (\ref{dr}) define a partial order
relationship, in the sense that given two densities $\rho,\rho'$, it may happen
that $\rho\nprec\rho'$ and $\rho'\nprec\rho$.

Let us now consider a general mixed state $\rho(\lambda)$ depending on a
continuous parameter $\lambda$. We will say that $\lambda$ is a {\it mixing
parameter} in a certain interval if $\rho(\lambda)$ becomes {\it more mixed} as
$\lambda$ increases in this interval:
\begin{equation}
\rho(\lambda)\prec\rho(\lambda')\;\;{\rm if}\;\;\lambda\geq \lambda'\,.
\label{rla}
\end{equation}
This is equivalent, in the case of a smooth dependence, to the condition
\[\partial s_j/\partial \lambda\leq 0\,,\;\; j=1,\ldots,n-1\,,\]
within this interval. The generalized entropy $S_f[\rho(\lambda)]$ is then a
{\it non-decreasing} function of $\lambda$ for {\it any} concave $f$, as easily
verified:
 \begin{equation}
 \frac{\partial S_f[\rho(\lambda)]}{\partial \lambda}=\sum_{j=1}^{n-1}
 \frac{\partial s_j}{\partial\lambda}[f'(p_j)-f'(p_{j+1})]\geq 0\,,\label{Sft}
 \end{equation}
since $f'(p_j)-f'(p_{j+1})\leq 0$ for $f$ concave. Such states exhibit then an
unambiguous disorder increase for increasing $\lambda$.

As a general example, let us consider the escort densities  \cite{BS.93}
 \begin{equation}
 \rho_q=\rho^q/Z_q,\;\;Z_q={\rm Tr}\,\rho^q\,, \label{sq1}
 \end{equation}
associated with a density matrix $\rho$. It is easily seen that for $q>0$,
$\lambda=1/q$ is a {\it mixing parameter} for $\rho_{q}$, since the ensuing
partial sums satisfy, for $j<n$,
 \begin{equation}
\frac{\partial s_j}{\partial\lambda}=-q^2\sum_{i=1}^j\sum_{k=j+1}^n
p_{qi}p_{qj}\ln(p_i/p_j)\leq 0,\;\;\;\lambda=1/q>0,
 \label{sq2}\end{equation}
where $p_{qi}=p_i^q/Z_q$ are the eigenvalues of $\rho_q$ ($p_{qi}\geq p_{qj}$
$\forall$ $q>0$ if $i\leq j$). Any other decreasing function of $q$ is of
course a mixing parameter for $\rho_q$ as well. We have therefore
$\rho_{q}\prec\rho_{q'}$ if $0<q\leq q'$ for a given fixed density matrix 
$\rho$.

Assume now that the system is described by a Hamiltonian $H$ with energies
$\varepsilon_i$, $i=1,\ldots,n$, sorted in what follows in {\it increasing}
order, and consider densities $\rho(\lambda)$ which satisfy the conditions:
$a)$ they {\it commute} with $H$, $b)$ their eigenvalues are {\it
non-increasing} functions of energy ($p_i\geq p_j$ if
$\varepsilon_i\leq\varepsilon_j$) and $c)$ $\lambda$ is a mixing parameter for
$\rho(\lambda)$ in a certain interval. In such a case, another fundamental
consequence of Eq.\ (\ref{rla}) is that not only the average energy $\langle
H\rangle_{\rho}={\rm Tr}\rho(\lambda)H$, but also the expectation value of {\it
any} non-decreasing function $w$ of $H$ ($w(\varepsilon_i)\leq
w(\varepsilon_j)$ if $i<j$), independent of $\lambda$, is a {\it
non-decreasing} function of $\lambda$:
\begin{equation}
\frac{\partial\langle w(H)\rangle_{\rho}}{\partial \lambda}=
\sum_{j=1}^{n-1}\frac{\partial s_j}{\partial \lambda}
[w(\varepsilon_j)-w(\varepsilon_{j+1})]\geq 0\,. \label{wt}
\end{equation}
This automatically ensures a non-negative generalized ``specific heat''
$c_\lambda\equiv\partial\langle H\rangle_{\rho}/\partial \lambda\geq 0$. Let us 
also remark that if $\partial s_j/\partial \lambda$ were positive for some $j$
(and $p_j>p_{j+1}$, $\varepsilon_j<\varepsilon_{j+1}$), one could always find
functions $f$ and $w$ of the previous forms such that Eqs.\ (\ref{Sft}) and
(\ref{wt}) become negative. In this way, one can in principle always witness
the absence of proper mixing increase.

We may now say that $\rho(\lambda)$ exhibits a {\it thermal-like behavior} if
in addition, it approaches the ground state density $I_1/n_1$ in some limit
$\lambda\rightarrow\lambda_0$ ($I_1$ denotes the projector onto the ground
state energy subspace and $n_1$ its degeneracy) and the state of maximum
disorder $I/n$ in some other limit $\lambda\rightarrow\lambda_\infty$, with
$\lambda$ a mixing parameter for $\lambda_0<\lambda<\lambda_\infty$. The most
common example of a state of the previous form is, of course, the standard
Boltzmann-Gibbs (BG) thermal state (we set in what follows Boltzmann constant 
$k=1$) 
 \begin{equation}
 \rho(T)=\exp[-H/T]/Z(T),\;\;Z(T)={\rm Tr}\exp[-H/T]\,,\;\;\;T>0\,, \label{BG}
 \end{equation}
obtained from the minimization of $\langle H\rangle_{\rho}-TS(\rho)$. It is
well known that its von Neumann entropy $S[\rho(T)]$ is an increasing function
of temperature, which is usually taken as the basis for the statement that
$\rho(T)$ becomes more disordered as $T$ increases. However, in the present
framework this statement can be more rigorously formulated. It is easily shown
that $T$ {\it is a proper mixing parameter} for $\rho(T)$ in the interval
$(0,\infty)$, i.e.,
\begin{equation}
\rho(T)\prec\rho(T')\;\;\;{\rm if}\;\;T\geq T'>0\,,
\end{equation}
as the sums of its first $j$ eigenvalues satisfy, for $j=1,\ldots,n-1$, 
\begin{equation}
\frac{\partial s_j}{\partial T}=\sum_{i=1}^j\sum_{k=j+1}^n
 p_ip_k(\varepsilon_i-\varepsilon_k)/T^2\leq 0\,.\label{sjtg}
\end{equation}
Hence, not only its von Neumann entropy, but also its {\it generalized entropy}
$S_f[\rho(T)]$, {\it is an increasing function of} $T$ for {\it any} concave
$f$. The average of {\it any} increasing function of energy is an increasing
function of $T$ as well. It can be shown that the generalized thermal density
obtained from the minimization of $\langle H\rangle_\rho-TS_f(\rho)$, given by
\cite{CR.02} $\rho_f(T)=f'^{-1}(H/T+\alpha)$, where $\alpha$ is a normalization
constant and the cutoff $f'^{-1}(u)=0$ if $u>f'(0)$ applies, also becomes more
mixed as $T$ increases \cite{RC.04}. Eq.\ (\ref{sjtg}) remains valid replacing
$p_ip_k$ by $\tilde{Z}\tilde{p}_i\tilde{p}_k$, with
$\tilde{p}_i=-[f''(p_i)\tilde{Z}]^{-1}$ for $p_i>0$ and
$\tilde{Z}=-\sum_{p_i>0}[f''(p_i)]^{-1}$, being both positive for $f$ concave.

Let us now identify the conditions which ensure that $\lambda$ is a mixing
parameter for a density of the more general form
\begin{equation}
\rho(\lambda)=g(H,\lambda)/Z(\lambda)\,,
\;\;Z(\lambda)={\rm Tr}\,g(H,\lambda)\,,\label{rhogl}
\end{equation}
where $g(\varepsilon,\lambda)$ is assumed to be an arbitrary smooth positive
non-increasing function of $\varepsilon$ for
$\varepsilon\in[\varepsilon_1,\varepsilon_n]$ depending on a parameter
$\lambda$, with $[\rho(\lambda),H]=0$. The variation rate of the associated
partial sums can be shown to be, for $j<n$,
 \begin{eqnarray}
 \frac{\partial s_j}{\partial \lambda}
&=&\sum_{i=1}^j\sum_{k=j+1}^n p_ip_k
[\tilde{g}_\lambda(\varepsilon_i)-\tilde{g}_\lambda(\varepsilon_k)]\,,\;\;\;
\tilde{g}_\lambda(\varepsilon)\equiv\frac{\partial\ln g}{\partial\lambda}
\,,\label{sjla}
 \end{eqnarray}
with $p_i=g(\varepsilon_i,\lambda)/Z(\lambda)$, which generalizes 
Eq.\ (\ref{sjtg}). A {\it sufficient} condition which ensures 
$\partial s_j/\partial\lambda\leq 0$ for $j=1,\ldots,n-1$ is, therefore, that 
$\tilde{g}_\lambda(\varepsilon)$ be a {\it non-decreasing} function of 
$\varepsilon$ for $\varepsilon\in[\varepsilon_1, \varepsilon_n]$, i.e.,
\begin{equation}
\frac{\partial^2\ln g}{\partial\varepsilon\partial\lambda}\geq 0 \;
\Rightarrow\;\frac{\partial s_j}{\partial \lambda}\leq
0\,,\;\;j=1,\ldots,n-1\,.\label{eq1}
\end{equation}
For instance, if
\begin{equation}
\rho(\lambda)=g(H/\lambda)/Z(\lambda)\,,\label{g2}
\end{equation}
where $\lambda>0$ and $g$ is here a positive non-increasing smooth real
function, $\tilde{g}_\lambda(\varepsilon)=-\lambda^{-1}u[\ln g(u)]'$, with
$u=\varepsilon/\lambda$, and Eq.\ (\ref{eq1}) leads to the condition
\begin{equation}
-\{u[\ln g(u)]'\}'\geq 0\,,\label{gx}
\end{equation}
which is not necessarily valid (consider for instance $g(u)=[1+\ln(u+1)]^{-1}$
for $u>0$). It is, of course, valid in the BG case for $\lambda=T$
($g(u)=\exp[-u]$) as well as for $g(u)=\exp(-u^r)$ $\forall$ $r>0$ if $u>0$
($-\{u[\ln g(u)]'\}'=r^2u^{r-1}\geq 0$). Note also that if $g(0)>0$ and
$g(\infty)=0$, the state $\rho(\lambda)=g(\bar{H}/\lambda)/Z(\lambda)$, with 
$\bar{H}=H-\varepsilon_1 I$, will always approach the ground state density
$I_1/n_1$ for $\lambda\rightarrow 0$ and the fully mixed state $I/n$ for
$\lambda\rightarrow\infty$. It is, however, the more stringent condition
(\ref{gx}) which ensures that $\rho(\lambda)$ will in addition become
monotonously {\it more mixed} as it evolves from the ground state to the fully
mixed state.

Note also that if $\lambda$ is a mixing parameter for the density
(\ref{rhogl}), it will remain a mixing parameter for the associated escort
density (\ref{sq1}) for $q>0$, which corresponds to
$g_q(H,\lambda)=g(H,\lambda)^q$. The sign of $\frac{\partial^2 \ln
g_q}{\partial\varepsilon\partial\lambda}$ is left unchanged for $\lambda$
independent of $q$. Eq.\ (\ref{sq2}) also follows from Eq.\ (\ref{sjla}) for
$g(H,\lambda)\rightarrow g(H)^{1/\lambda}$.

As an important example of Eq.\ (\ref{rhogl}), we will examine the mixing
properties of density operators characterized by a power-law distribution,
which can be written in the form of the  Tsallis distribution
\cite{TS.88,Cu.91}
\begin{equation}
\rho(q,T^*)=[I-(1-q)\bar{H}/T^*]_+^{\frac{1}{1-q}}/Z(q,T^*)\,,\label{rqt}
\end{equation}
where $T^*>0$ represents an effective temperature, $q$ the non-extensivity
index, $\bar{H}=H-\varepsilon_1 I$ the energy measured from the ground state
and $[u]_+\equiv (u+|u|)/2$. For $q\rightarrow 1$, $\rho(q,T)$ approaches the 
BG distribution (\ref{BG}). Eq.\ (\ref{rqt}) is 
obviously positive and fulfills previous conditions $a)$ and $b)$ $\forall$
$T^*>0$ and $q\in\Re$. Its eigenvalues $p_i$ are strictly decreasing functions
of energy for $q>1$, but just non-increasing functions for $q<1$ due to the
{\it cutoff} that applies in this case ($p_i=0$ if
$(1-q)\bar{\varepsilon}_i\geq T^*$, where $\bar{\varepsilon}_i\equiv
\varepsilon_i-\varepsilon_1$).

We will now show that {\it both} $T^*$ and $q$ are {\it proper mixing
parameters} for $\rho(q,T^*)$. Defining $g(\bar{\varepsilon},q,T^*)\equiv
[1-(1-q)\bar{\varepsilon}/T^*]_+^{\frac{1}{1-q}}$, we obtain
\begin{eqnarray}
\frac{\partial^2\ln g}{\partial\bar{\varepsilon}\partial T^*}
&=&\frac{1}{(T^*-(1-q)\bar{\varepsilon})^2}\geq 0\,,\;\;
\frac{\partial^2\ln g}{\partial\bar{\varepsilon}\partial q}=
\frac{\bar{\varepsilon}}{(T^*-(1-q)\bar{\varepsilon})^2}\geq 0\,,
\end{eqnarray}
for $\bar{\varepsilon}>0$ and $(1-q)\bar{\varepsilon}/T^*<1$, so that
according to Eq.\ (\ref{eq1}), $\rho(q,T^*)$ becomes {\it more mixed as either
$T^*$ or $q$ increases}: 
 \begin{eqnarray}
 \rho(q,T^*)&\prec &\rho(q,{T'}^*)\;{\rm if}\; T^*\geq {T'}^*>0\,,\;\;\;
 \rho(q,T^*)\prec\rho(q',T^*)\;{\rm if}\; q\geq q'\,. \label{mx2}
\end{eqnarray}
The role of $T^*$ and $q$ as proper mixing parameters constitutes then another
fundamental property of the distribution (\ref{rqt}). {\it Any} entropy
$S_f[\rho(q,T^*)]$ (in particular $S_{q'}[\rho(q,T^*)]$ for {\it any} $q'>0$)
is a non-decreasing function of both $T^*$ {\it and} $q$ in  ${\it any}$
system, as illustrated in fig.\ 1 for a truncated harmonic oscillator. 

\begin{figure}
\vspace{0cm}
\hspace*{0cm}\centerline{\scalebox{1.}{\includegraphics{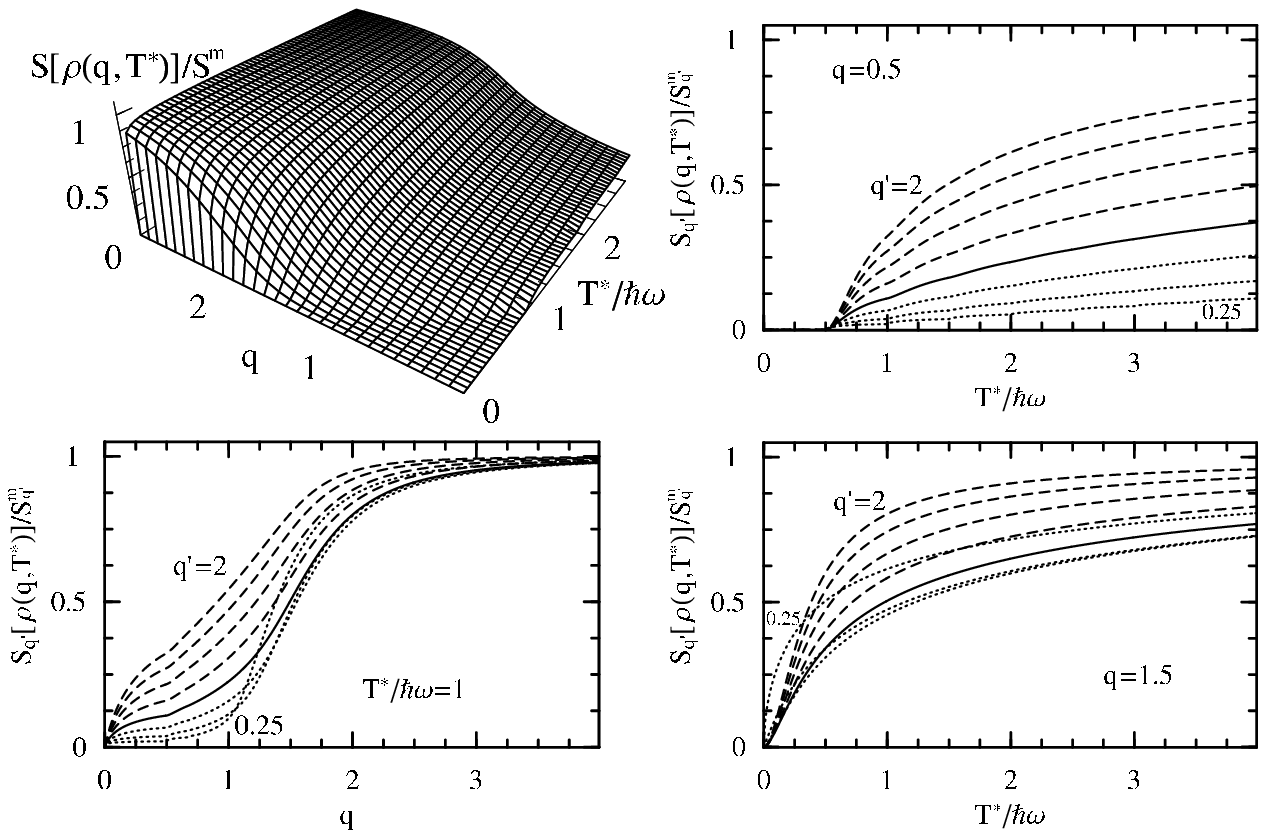}}}
%\vspace*{-19.7cm}
\vspace*{-0.cm}

\caption{Top left: The von Neumann entropy of the Tsallis distribution
(\ref{rqt}) (scaled to the maximum value $S^m\equiv S(I/n)$) for $n=100$
equally spaced levels with spacing $\hbar\omega$, showing that it is a
non-decreasing function of both $T^*$ and $q$. Remaining panels: The scaled
Tsallis entropy $S_{q'}[\rho(q,T^*)]/S_{q'}^m$ in the same system, as a 
function of $T^*$ at fixed $q$ (right panels) and as a function of $q$ at fixed 
$T^*$ (bottom left panel), for different values of $q'$, ranging from $q'=0.25$  
to $q'=2$ in steps of $0.25$. Dashed (dotted) lines correspond to $q'>1$ 
($q'<1$), solid lines to $q'=1$ (von Neumann entropy). They are all increasing 
functions of $T^*$ and $q$.}\label{f1}\end{figure}

Let us remark that Eq.\ (\ref{rqt}) exhibits {\it a proper thermal-like
behavior with respect to both $q$ and $T^*$}, since in addition it approaches
the ground state density $I_1/n_1$ both for $T^*\rightarrow 0$ at fixed $q$ and
for $q\rightarrow -\infty$ at fixed $T^*$, and the random state $I/n$ both for
$T^*\rightarrow\infty$ at fixed $q$ as well as for $q\rightarrow\infty$ at
fixed $T^*$. Actually, due to the cutoff for $q<1$, $\rho(q,T^*)=I_1/n_1$
already for $T^*<(1-q)\Delta$ at fixed $q<1$, and $q<1-T^*/\Delta$ at fixed
$T^*$, where $\Delta$ is the lowest non-zero excitation energy. Note also that
$\rho(q,T^*)-I/n$ behaves {\it linearly} with $\bar{H}$ for large $T^*$
($g(\bar{\varepsilon},q,T^*)\approx 1-\bar{\varepsilon}/T^*$ for $T^*\gg
|1-q|\bar{\varepsilon}$\,) but {\it logarithmically} for large $q$
($g(\bar{\varepsilon},q,T^*)\approx 1-\ln[1+q\bar{\varepsilon}/T^*]/q$ for
$q\gg\ln[1+q\bar{\varepsilon}/T^*]$ and $q\gg 1$).

Eq.\ (\ref{rqt}) can also be characterized by other mixing parameters. For
example, rewriting Eq.\ (\ref{rqt}) as $\rho_s(\gamma,\mu)=
~[1-s\bar{H}/\mu]_+^{s/\gamma}/Z(\gamma,\mu)$, with $\gamma=|1-q|>0$, 
$\mu=T^*/|1-q|>0$ and $s={\rm Sign}(1-q)$, {\it both} $\mu$ and $\gamma$ are as
well independent mixing parameters for both signs of $s$ (despite the increase
of $\gamma$ with decreasing $q$ for $q<1$), since
 $\frac{\partial^2\ln g}{\partial\bar{\varepsilon}\partial\mu}=
 [\gamma(\mu-s\bar{\varepsilon})^2]^{-1}\geq0$,
 $\frac{\partial^2\ln g}{\partial\bar{\varepsilon}\partial\gamma}=
 [\gamma^2(\mu-s\bar{\varepsilon})]^{-1}\geq 0$ for $s\bar{\varepsilon}/\mu<1$.
Thus, $\rho_s(\gamma,\mu)\prec\rho_s(\gamma,\mu')$ if $\mu\geq\mu'$ and
$\rho_s(\gamma,\mu)\prec\rho_s(\gamma',\mu)$ if $\gamma\geq\gamma'$. These
properties also follow from Eq.\ (\ref{gx}) in the case of $\mu$ (for
$g(u)=[1-su]_+^{s/\gamma}$) and from Eqs.\ (\ref{sq1})--(\ref{sq2}) in the case
of $\gamma$. Again, we obtain a thermal-like behavior with respect to both
$\gamma$ and $\mu$ for $s=\pm 1$, with the ground state density approached for
$\mu\rightarrow 0$ or $\gamma\rightarrow 0$ and the full random state
approached for $\mu\rightarrow\infty$ or $\gamma\rightarrow\infty$ (provided
$\mu>\bar{\varepsilon}_n$ if $s=1$).

The actual thermal state derived from the non-extensive thermodynamic formalism
based on the Tsallis entropy and the minimization of the free energy
$F_q=\langle H\rangle_{\rho_q}-TS_q(\rho)$ \cite{TM.98,TS.04}, where $q>0$ and
$\rho_q$ is the escort density (\ref{sq1}), is also of the form (\ref{rqt}) but
with $T^*$ related to the actual $T$ by
$T=[T^*-(1-q)\langle\bar{H}\rangle_{\rho_q}]/Z_q$ (following ref.\
\cite{TS.04}). For $q>1$, $T$ is a direct increasing function of $T^*$, as in
this case $\partial T/\partial T^*=(\eta q-1)/[(q-1)Z_q]\geq 0$ (with
$\eta=2\langle O\rangle_{\rho_q}\langle
O^{-1}\rangle_{\rho_q}-1\geq 1$ and $O=I-(1-q)\bar{H}/T^*$), so that $T$
will also be a {\it proper mixing parameter}. It will remain so for $0<q<1$
provided the absolute minimum of the free energy $F_q$ at each $T$ is
considered \cite{LP.99}, as in this case the entropy (and hence $T^*$) cannot
decrease with increasing $T$.

So far all previous expressions are applicable in both the quantum and the
classical discrete case. Let us finally briefly examine the majorization
properties of density matrices constructed from constraints on two or more {\it
non-commuting observables.} We may for instance consider two observables $H_0$,
$H_1$, with $[H_0,H_1]\neq 0$, and a density of the form
\begin{equation}\rho(\lambda_0,\lambda_1)=g(H_0/\lambda_0+H_1/\lambda_1)
/Z(\lambda_0,\lambda_1)\,,\label{rho2}
\end{equation}
where $g(u)$ is a positive non-increasing function and $\lambda_\nu>0$ for 
$\nu=0,1$, generalizing Eq.\ (\ref{g2}).  In this case, Eq.\ (\ref{sjla}) 
should be replaced, for $j<n$, by \vspace*{-.25cm}
\begin{eqnarray}
 \frac{\partial s_j}{\partial \lambda_\nu}
&=&\sum_{i=1}^j\sum_{k=j+1}^n p_ip_k
[\tilde{g}^\nu_{i}-\tilde{g}^\nu_{k}]\,,\;\;\;
\tilde{g}^\nu_{i}=\frac{\partial \ln g(u_i)}{\partial \lambda_\nu}=
-\frac{g'(u_i)}{\lambda_\nu^2g(u_i)}\langle i|H_\nu|i\rangle\,,\label{go2}
 \end{eqnarray}
where $u_i$ denotes the eigenvalues (sorted in increasing order) of
$\sum_{\nu=0,1} H_\nu/\lambda_\nu$ and $|i\rangle$ the corresponding
eigenstates (in case of degeneracy we assume in (\ref{go2}) $H_\nu$ diagonal
within each eigenspace). Hence, we can ensure that $\lambda_\nu$ will be a
mixing parameter for (\ref{rho2}) if  $\tilde{g}^\nu_{i}$ {\it does not
decrease for increasing values of $u_i$}.

For a single observable $H_0$, $\tilde{g}^0_i=-\lambda_0^{-1}u_ig'(u_i)/g(u_i)$ 
and the previous condition reduces to Eq.\ (\ref{gx}). This is also the case
when $\langle i|H_\nu|i\rangle$ is {\it proportional} to $u_i$ ($\langle
i|H_\nu|i\rangle=\alpha u_i$, with $\alpha>0$ and independent of $i$), as
occurs in simple systems such as a harmonic oscillator $H=[p^2/m+kx^2]/2$. In
this case $\langle i|p^2|i\rangle=mE_i$, $\langle i|x^2|i\rangle=E_i/k$, with
$E_i=\hbar\omega(i+1/2)$ the oscillator energies ($\omega=\sqrt{k/m}$), so that
both the {\it mass} $m$ and the {\it inverse oscillator strength} $k^{-1}$ are
also {\it mixing parameters} in $\rho=g(H/T)/Z(T)$, provided $T$ is a mixing
parameter too (for $\lambda_0=m$, $\lambda_1=k^{-1}$, we would have
$g_{i}^\nu=-\frac{1}{2}\lambda_\nu^{-1}[u_ig'(u_i)/g(u_i)]$, with $u_i=E_i/T$).
In the general case, however, $\langle i|H_\nu|i\rangle/u_i$ may depend on $i$
in a non-trivial way, so that the mixing properties of $\rho$ will require a
careful analysis of the behavior of $\tilde{g}^\nu_i$.

In conclusion, we have applied the theory of majorization to identify the
rigorous sufficient mixing conditions, as well as their main physical
implications, for general mixed states of the form (\ref{rhogl}), which are
summarized in Eqs.\ (\ref{eq1}), (\ref{gx}), (\ref{Sft}) and (\ref{wt}). As 
application, we have examined those characterized by a power law distribution, 
and shown that they can be expressed in terms of two fundamental mixing 
parameters, which can be taken as $T^*$ and $q$ in the representation 
(\ref{rqt}). We have in particular identified the role of $q$ in (\ref{rqt}) as 
a rigorous mixing parameter. The actual thermal state derived in the Tsallis 
non-extensive thermodynamic formalism was also shown to become more mixed for 
increasing $T$ (with the above remarks applying for $0<q<1$), as occurs with 
the standard BG thermal state, ensuring in particular a universal entropy 
increase, i.e., $\partial S_f[\rho(q,T)]/\partial T\geq 0$ for {\it any} 
concave $f$. These results strengthen thus the robustness of the generalized 
non-extensive thermodynamic formalism. We have also discussed the majorization 
properties of escort distributions [Eq.\ (\ref{sq1})] and derived sufficient 
conditions for mixedness increase in the presence of non-commuting observables 
[Eq.\ (\ref{go2})].

Majorization theory enables then to derive very general inequalities with deep
implications by simple means. Generalized thermal-like distributions with
proper mixing parameters may also help to provide a more complete description
of the behavior of a correlated quantum system with increasing randomness,
revealing aspects which could be hidden in standard BG statistics
\cite{RC.04}. For instance, any system possessing a limit temperature $T_c$ in
standard statistics for some property present in its ground state and absent in
the vicinity of the completely random state, will also possess a limit value of
the mixing parameter in a generalized thermal-like distribution. Knowledge of
such boundaries (like a critical curve $T^*_c(q)$ in (\ref{rqt}))
may provide a new perspective for the classification of order-disorder
transitions or crossovers.
\\The authors acknowledge support from CIC (RR), CONICET (NC,MP) and 
ANPCYT (MP) of Argentina.  \vspace*{-.25cm}

%%%%%%%%%%%%%%%%%%%%%%%%%%%%%%%%%%%%%%%%%%%%%

\end{document}